\documentclass[12pt]{article}

\usepackage{amsmath}
\usepackage{amssymb}
\usepackage{epsfig}
\usepackage{psfrag}
\usepackage{cite}

\def\no{\nonumber}
\def\calC{\mathcal{C}}
\def\calA{\mathcal{A}}
\def\MSbar{$\overline{\text{MS}}$}

\allowdisplaybreaks[1]

\begin{document}


\begin{titlepage}

\begin{flushright}
 TTP10-22
\end{flushright}
\vskip 2.6cm

\begin{center}
\Large{\bf\boldmath
Electroweak Sudakov Logarithms and  \\
Real Gauge-Boson Radiation in the TeV Region
\unboldmath}

\normalsize
\vskip 1.5cm

{\sc G.~Bell}$^{a,b}$,
{\sc J.~H.~K\"uhn}$^{a}$
and
{\sc J.~Rittinger}$^{a}$

\vskip .5cm

{\it $^a$
Institut f\"ur Theoretische Teilchenphysik, \\
Universit\"at Karlsruhe, D-76128 Karlsruhe, Germany}\\[0.3cm]
{\it $^b$
Albert Einstein Center for Fundamental Physics,\\
Institute for Theoretical Physics,
University of Bern,\\
CH-3012 Bern,
Switzerland}

\vskip 2.4cm

\end{center}

\begin{abstract}
\noindent
Electroweak radiative corrections give rise to large negative,
double-logarithmically enhanced corrections in the TeV region. These are
partly compensated by real radiation and, moreover, affected by
selecting isospin-noninvariant external states. We investigate the
impact of real gauge boson radiation more quantitatively by considering
different restricted final state configurations. We consider
successively a massive abelian gauge theory, a spontaneously broken
$SU(2)$ theory and the electroweak Standard Model. We find that details
of the choice of the phase space cuts, in particular whether a fraction
of collinear \emph{and} soft radiation is included, have a strong impact
on the relative amount of real and virtual corrections. 
\end{abstract}

\vfill

\end{titlepage}


\section{Introduction}


During the past years electroweak radiative corrections have been
evaluated for numerous lepton and hadron collider processes. Despite the
relatively small coupling, $\alpha_W/\pi = \alpha/(\pi\sin^2
\theta_w)\simeq 0.01$, virtual gauge boson exchange becomes important at
high energies, a consequence of the enhancement by large ''Sudakov''
logarithms~\cite{Sud,Jac} with the dominant terms proportional to $\ln^2
s/M_{W,Z}^2$. 

In contrast to QED and QCD, where physical cross sections are obtained
by combining virtual and real radiation, events with and without real
$W$- and $Z$-bosons have a distinctly different signature and as such
they can in principle be separated in an experimental setup. This
observation has led to numerous studies for exclusive
reactions~\cite{Kur,DegSir,Bec,CiaCom,KuhPen,Fad,KPS,KMPS,Bec2,BRV,Mel,DenPoz,HKK,BeeWet,FKM,DMP,Bec3,Bec4,Poz,FKPS,JKPS,Jan,KMP,Man},
ranging from purely electroweak four-fermion processes or $W$-pair
production at electron-positron colliders to the hadronic production of
$Z$, $\gamma$~\cite{MMR,KKPS1} or $W$~\cite{KKPS2,HKK1} at large
transverse momenta. In some of these cases, in particular for scattering
energies in the TeV region, the (negative) one-loop corrections amount
to $10\%$ or even up to $30\%$. This has motivated the investigation of
logarithmically enhanced terms of higher orders, either from two-loop
effects or in a resummed all-order formulation. To obtain the leading
logarithmic (LL) and next-to-leading logarithmic (NLL) terms is
straightforward, however, at the same time insufficient for an adequate
description e.g.~of the dominant two-loop terms. This has motivated
studies of higher order contributions and NNLL, partly even NNNLL,
results are available for many reactions~\cite{KPS,KMPS,FKPS,JKPS,KMP}.

The crucial assumption in most of these studies, that events with real
gauge boson radiation can be discriminated from the ''exclusive'' final
state, has to be justified by a detailed analysis which obviously
depends on the experimental setup. In particular significant differences
are expected between electron-positron and hadron colliders, and between
leptons or quark and gluon jets in the final state. For ''clean''
reactions like lepton- or gauge boson-pair production in
electron-positron collisions one may anticipate a clear separation, for
quark jets in the final state at an hadron collider like the LHC the
situation is expected to be more involved.   

This has motivated a detailled study of weak boson emission at hadron
colliders~\cite{Baur:2006sn}, which demonstrates that, although partial
cancellations between virtual and real radiation may occur, the real
emission process often only compensates part of the virtual
corrections. At first glance one might expect that the combination of
virtual and real radiation, the latter completely inclusive, would lead
to a complete compensation of the Sudakov logarithms. However, as
observed in~\cite{CiaCiaCom}, the preparation of isospin non-invariant
external states like electrons or up and down quarks at
electron-positron or hadron colliders, respectively, leads to a
non-vanishing logarithmically enhanced remainder, a phenomenon called
Bloch-Nordsieck violations. These studies were performed in the
high-energy limit and real radiation was treated in a completely
inclusive manner.  

In the present paper we investigate the relative size of virtual versus
real radiation, imposing a variety of cuts on the phase space of the
emitted gauge boson. These cuts are supposed to represent, in somewhat
idealized form, constraints arising from typical detector
configurations. As characteristic examples we will consider final states
with soft gauge bosons or, alternatively, with gauge bosons collinear to
incoming or outgoing particles. Our considerations will allow to
''interpolate'' between the completely exclusive and the inclusive
treatments. Furthermore, for simplicity, the discussion will be limited
to four-fermion processes.  

The outline of this paper is as follows. In Section~\ref{sec:cuts} we
work out the generic structure of Sudakov logarithms in real emission
processes with phase space cuts. While this discussion will be limited
to an abelian gauge theory, we investigate the structure and the
numerical impact of the Bloch-Nordsieck violations in a spontaneously
broken $SU(2)$ theory in Section~\ref{sec:BN}. The size of the
Bloch-Nordsieck violations will be compared to the difference between
the fully inclusive result and the one with restricted phase space. In
Section~\ref{sec:SM} our predictions for the Standard Model are
presented. As a working example we consider the process $e^+e^-\to q
\bar{q}$. We compute next-to-leading order (NLO) electroweak corrections
and, in particular, investigate the compensation of the Sudakov
suppression from unobservable $W$- and $Z$-boson radiation. We finally
conclude in Section~\ref{sec:conclusion}.


\section{Real emission with phase space restrictions}

\label{sec:cuts}

We first pursue the question to which extent the virtual Sudakov
corrections are compensated by the real emission process if the latter
is subject to certain phase space restrictions. In contrast to the
familiar picture from QED or QCD, where an infrared-safe observable
necessarily requires inclusion of soft \emph{and} collinear gauge boson
emission, it will be instructive for our purposes to also consider
scenarios that allow for soft \emph{or} collinear radiation. The
physical relevance of the particular phase space cuts will depend on the
details of the observable under consideration.  We therefore relegate
this question to our phenomenological analysis in Section~\ref{sec:SM}
and concentrate for the moment on the generic structure of Sudakov
logarithms in real emission processes with phase space restrictions.   

It will be convenient for the current discussion to work in a first step
within a toy theory that captures the physics of interest while allowing
for a compact and transparent presentation. To be specific we consider
an abelian gauge theory with explicit mass term $\frac12 M^2 A_\mu
A^\mu$ that spoils the gauge invariance, but leads, nevertheless, to a
consistent renormalizable theory. In the remainder of this section we
first address Sudakov logarithms that arise from final state radiation,
subsequently we generalize the discussion to the four-fermion process.

\subsection{Final state radiation}

Let us start with an elementary process, namely with the decay of a
heavy vector boson (with mass $\sqrt{s}$) into a pair of massless
fermions.  We assume that this initial vector boson does not couple to
our toy theory and that the decay is mediated at Born level by some
other vectorlike interaction which we do not specify further. The
one-loop virtual corrections are then entirely encoded in the abelian
vector form factor (in the timelike region), which has been the central
object in the study of electroweak Sudakov logarithms. In the Sudakov
limit $s\gg M^2$, the explicit one-loop calculation yields  
\begin{align}
\Gamma^{(V)}
&\simeq \frac{\alpha}{4\pi}
\left \{ - 2 \left[ \ln^2 \frac{s}{M^2} - 3 \ln \frac{s}{M^2} \right] 
+ \frac{2\pi^2}{3} - 7 \right\} \Gamma_B,
\label{eq:U1:decay:virtual}
\end{align}
where the fermion wave functions have been renormalized in the on-shell
scheme and $\Gamma_B$ is the Born decay rate. The result reveals a
characteristic structure that dominates the decay rate in the
high-energy limit. It contains a \emph{Sudakov factor}
$\alpha/(4\pi)\,[\ln^2 s/M^2 - 3 \ln s/M^2]$ with negative weight (and
proportional to the charge squared) for each of the interacting
fermions. We will see in the following section that the Sudakov factor
is process-independent and contains the full information about collinear
logarithms.  

We next consider the corresponding real emission process, where the
light vector boson is emitted from the final state fermions. Without any
restriction on the phase space of the emitted boson, the decay rate
becomes in the Sudakov limit  
\begin{align}
\Gamma^{(R)}
&\simeq \frac{\alpha}{4\pi}
\left \{ 2 \left[ \ln^2 \frac{s}{M^2} - 3 \ln \frac{s}{M^2} \right]
- \frac{2\pi^2}{3} + 10 \right\} \Gamma_B.
\label{eq:U1:decay:real:inclusive}
\end{align}
We see that the logarithmic terms cancel in the sum of virtual and real
corrections, in accordance with the expectations from the
Kinoshita-Lee-Nauenberg (KLN) theorem~\cite{Kin,LeeNau}.  

Let us now examine how the pattern of these logarithms changes, when we
impose different restrictions on the phase space of the emitted
boson. We first consider a scenario that allows for soft \emph{and}
collinear radiation. To this end we require that the final state
fermions are almost back-to-back in the center of mass frame, i.e.~we
impose a cutoff on the opening angle of the fermion pair, $\theta_{f\bar
  f}\geq\theta_{f\bar f}^c$, with $\theta_{f\bar f}^c$ close to
$180^\circ$. In other words we only exclude hard and non-collinear
radiation. We now obtain for the restricted real emission process (with
$c^c_{f\bar f}\equiv\cos\theta_{f\bar f}^c$)  
\begin{align}
\Gamma^{(R)}(\theta_{f\bar f}^c)
&\simeq \frac{\alpha}{4\pi}
\left \{ 2 \left[ \ln^2 \frac{s}{M^2} - 3 \ln \frac{s}{M^2} \right]
+ 4 \text{Li}_2\left(-\frac{1-c^c_{f\bar f}}{1+c^c_{f\bar f}}\right)
\right. 
\no\\
& \hspace{14mm} \left.
+ \frac{8(2-c^c_{f\bar f})}{(1-c^c_{f\bar f})^2}\,
  \ln \left(\frac{2}{1+c^c_{f\bar f}} \right)
+ \frac{1-5c^c_{f\bar f}}{1-c^c_{f\bar f}}
- \frac{2\pi^2}{3} \right\} \Gamma_B,
\end{align}
which holds for $s\gg M^2$ and $1+c^c_{f\bar f}\gg M^2/s$. We see that
the given phase space cut does not modify the structure of the mass
singularities at all. As the restricted phase space covers all of the
singular regions, we again obtain the full Sudakov factors and hence
observe a complete cancellation between virtual and real Sudakov
logarithms.  

It is also interesting to consider a highly restricted phase space in
the given scenario, which corresponds to the limit $\theta_{f\bar
f}^c\to180^\circ$. We then find   
\begin{align}
\Gamma^{(R)}(\theta_{f\bar f}^c\to180^\circ)
&\simeq \frac{\alpha}{4\pi}
\left \{ 2 \left[ \ln^2 \frac{s}{M^2} - 3 \ln \frac{s}{M^2} \right]
\right. 
\no\\
& \hspace{14mm} \left.
-2 \left[ \ln^2 \left( \frac{2}{1+c^c_{f\bar f}} \right)
- 3  \ln \left( \frac{2}{1+c^c_{f\bar f}} \right) \right]
-\frac{4\pi^2}{3} + 3 \right\} \Gamma_B,
\end{align}
which illustrates how mass singularities are translated into phase space
logarithms if only a small fraction of soft and collinear radiation is
taken into account. The question of whether or not the Sudakov
logarithms numerically dominate the decay  rate finally depends on the
size of the argument in the phase space logarithms. From our explicit
result we are led to expect that the virtual Sudakov logarithms are
largely compensated if only a loose cutoff on the opening angle of the
fermion pair is applied with $\theta_{f\bar f}^c\lesssim 160^\circ$.  

We next consider a different scenario that allows for collinear
radiation only. We now require that the emitted boson is almost parallel
to one of the final state fermions, i.e.~we impose the constraints
$\theta_{f b}\leq\theta_{F b}^c$ or $\theta_{\bar f b}\leq\theta_{F
b}^c$ on the angles between the emitted vector boson and the outgoing
fermions. Let us now focus for simplicity on the singular part, which is
found to be (with $c_{Fb}^c\equiv\cos\theta_{F b}^c$)  
\begin{align}
\Gamma^{(R)}(\theta_{F b}^c)
&\simeq \frac{\alpha}{4\pi}
\left \{ 2 \left[ \ln^2 \frac{s}{M^2} - 3 \ln \frac{s}{M^2} \right]
- 4 \ln \left(\frac{1+c_{Fb}^c}{1-c_{Fb}^c}\right) \ln \frac{s}{M^2} 
+ \ldots  \right\} \Gamma_B,
\end{align}
which holds for $s\gg M^2$ and $1\geq1-c_{Fb}^c\gg M^2/s$ (we are
actually only interested in the region $1\gg1-c_{Fb}^c\gg
M^2/s$). Whereas the double logarithms again cancel between virtual and
real corrections, the linear logarithms do not (for $\theta_{F
b}^c<90^\circ$). The incomplete cancellation reflects the fact that the
considered scenario does \emph{not} cover all of the singular regions,
it misses in particular soft radiation that escapes the two cones around
the final state fermions. We thus expect that the compensation of the
virtual Sudakov logarithms is again significant but less effective in
this scenario.  

For completeness let us also consider a scenario that allows for soft
radiation only. We now impose a cutoff on the momentum of the vector
boson $|\vec{k}|\leq k_c$ or, equivalently, on the invariant mass of the
fermion pair $Q^2\geq Q_c^2$. As long as we do not cut into the endpoint
region, i.e.~for $k_c\gg M$ or $s-Q_c^2\gg 2M\sqrt{s}$, we obtain (with
$z_c\equiv Q_c^2/s$)  
\begin{align}
\Gamma^{(R)}(z_c)
&\simeq \frac{\alpha}{4\pi}
\left \{ 2 \left[ \ln^2 \frac{s}{M^2} - 3 \ln \frac{s}{M^2} \right]
+ 2 \Big[ 4 \ln (1-z_c) + 2z_c + z_c^2 \Big] \ln \frac{s}{M^2}
+ \ldots  \right\} \Gamma_B.
\label{eq:U1:decay:real:soft}
\end{align}
We thus find a situation that is conceptually similar to the one
before. We again observe an incomplete cancellation of the linear
logarithms (for $z_c>0$) since part of the singular region from
(hard-)collinear emission is missed.

\subsection{Four-fermion process}
\label{sec:U1:fourfermion}

The preceding example allowed us to study the generic structure of
Sudakov logarithms in real emission processes with some exemplary (and
idealized) phase space restrictions. Before making any quantitative
statements, let us now switch to the four-fermion process which brings
in two new aspects. First, we have to deal with initial state radiation
and, second, we have to consider the interplay of several phase space
restrictions.   

The one-loop virtual corrections to the $s$-channel four-fermion process
$f'\bar f' \to f \bar{f}$ amount to the calculation of two form factor
type corrections, two box diagrams and the vacuum polarization. As the
interference between tree and box diagrams vanishes in our abelian toy
theory, the result takes a particularly simple form 
\begin{align}  
\sigma^{(V)}
&\simeq \frac{\alpha}{4\pi}
\left \{ - 4 \bigg[ \ln^2 \frac{s}{M^2} - 3 \ln \frac{s}{M^2} \bigg] 
+ \frac{4\pi^2}{3} - 14
-\frac{40}{9} n_f -\frac{16}{9} n_s
\right\} \sigma_B,
\label{eq:U1:fourfermion:V}
\end{align}
where $\sigma_B\simeq 4\pi\alpha^2/3s$ is the Born cross section. We
thus obtain twice the form factor correction (\ref{eq:U1:decay:virtual})
and a contribution from the vacuum polarization\footnote{We assume that
the theory contains $n_f$ massless fermions and $n_s$ (light) scalar
bosons and renormalize the coupling constant in the \MSbar-scheme. We
further set the renormalization scale $\mu=\sqrt{s}$.}.  

In the corresponding real emission process the vector boson can be
emitted from initial and  final state fermions. In contrast to the
previous example, we now impose a cutoff on the invariant mass of the
fermion pair from the beginning, $Q^2\geq Q_c^2\gg M^2$, which allows us
to circumvent the $s$-channel resonant contribution from the initial
state radiation (which is of no particular interest for us since we
focus on final state configurations that resemble the four-fermion
process). Without further restrictions on the emission process, the
cross section now becomes in the logarithmic approximation (with
$z_c\equiv Q_c^2/s$)  
\begin{align}
\sigma^{(R)}(z_c) &\simeq
\frac{\alpha}{4\pi}
\left\{ 4 \bigg[ \ln^2 \frac{s}{M^2} - 3 \ln \frac{s}{M^2} \bigg]
\right. 
\no\\
&\hspace{14mm}\left.
+ 2 \left[ 1 + 4z_c + z_c^2 - 2\ln \frac{ z_c  }{ (1-z_c)^4} \right] \ln
\frac{s}{M^2} 
+ \ldots \right\} \sigma_B,
\label{eq:U1:fourferm:real:Q2}
\end{align}
where we assumed that $M^2\ll Q_c^2 \ll s-2M\sqrt{s}$. Due to the
explicit cutoff $Q_c^2$, we thus start with a mismatch between linear
virtual and real logarithms from the beginning.  

Even if we considered a fully inclusive observable, we actually would
not expect the logarithms from initial state radiation to completely
cancel the corresponding virtual ones. This may be illustrated with the
differential cross section in $z\equiv Q^2/s$,  
\begin{align}
\frac{d\sigma^{(R)}}{dz} &\simeq
\frac{\alpha}{4\pi} \left\{
\frac{1}{z}
\left(4\, \frac{1+z^2}{1-z}
\ln\frac{(1-z)^2 s}{z M^2}
-8 (1-z) \right) \right.
\no\\
&\hspace{14mm}
\left.
+ \left( 4\, \frac{1+z^2}{1-z}
\ln\frac{(1-z)^2 s}{M^2}
-8 (1-z) \right) \right\} \sigma_B,
\label{eq:U1:fourferm:real:diff}
\end{align}
where the first (second) line contains the initial (final) state
radiation\footnote{The interference between initial and final state
radiation vanishes in the abelian toy theory in analogy to the
cancellation of the box diagrams mentioned above.}. Integrating the
second line in the kinematic limits $0\leq z \leq (1- M/\sqrt{s})^2$, we
recover the Sudakov logarithms from the inclusive final state radiation
in~(\ref{eq:U1:decay:real:inclusive}). For the initial state radiation,
however, we have to proceed differently to single out the logarithms
that match the according virtual ones. Applying the usual prescription
for plus-distributions, we get  
\begin{align}
\sigma^{(R)}_\text{initial} &\simeq
\frac{\alpha}{4\pi}
\left\{ 2 \bigg[ \ln^2 \frac{s}{M^2} - 3 \ln \frac{s}{M^2} \bigg]
+ 4 \int_0^1 \frac{dz}{z} \left[ \frac{1+z^2}{1-z}  \ln
  \frac{(1-z)^2s}{M^2} \right]_+ 
+ \ldots \right\} \sigma_B,
\label{eq:U1:fourferm:real:plusdis}
\end{align}
which illustrates that there is a single collinear logarithm left that
does not cancel between virtual and real corrections. The reason for
this incomplete cancellation is of course well-known; according to the
KLN theorem we would have to account for \emph{incoming} vector bosons
to recover complete cancellation. In QCD applications the remnant
collinear singularity is usually factorized into process-independent
parton distribution functions and similar methods are used in the
context of QED. For the weak interactions with a physical gauge boson
mass, however, there is no need to factorize this contribution and one
is left with a certain mismatch in a fixed-order calculation.  

\begin{figure}[t!]
\centerline{\parbox{15cm}{
\centerline{\small
 \psfrag{A}{\hspace{-5.5mm}\underline{Scenario~A:}}
 \psfrag{B}{\hspace{-5.5mm}\underline{Scenario~B:}}
 \psfrag{fbar}{~$\bar f$}
 \psfrag{f}{\hspace{-2mm}$f$}
 \psfrag{fbarp}{$\bar f'$}
 \psfrag{fp}{$f'$}
 \psfrag{thetaI}{$\theta_{Ib}^c$}
 \psfrag{thetaF}{$\theta_{Fb}^c$}
 \psfrag{thetaff}{$\theta_{f\bar f}^c$}
 \includegraphics[width=7cm]{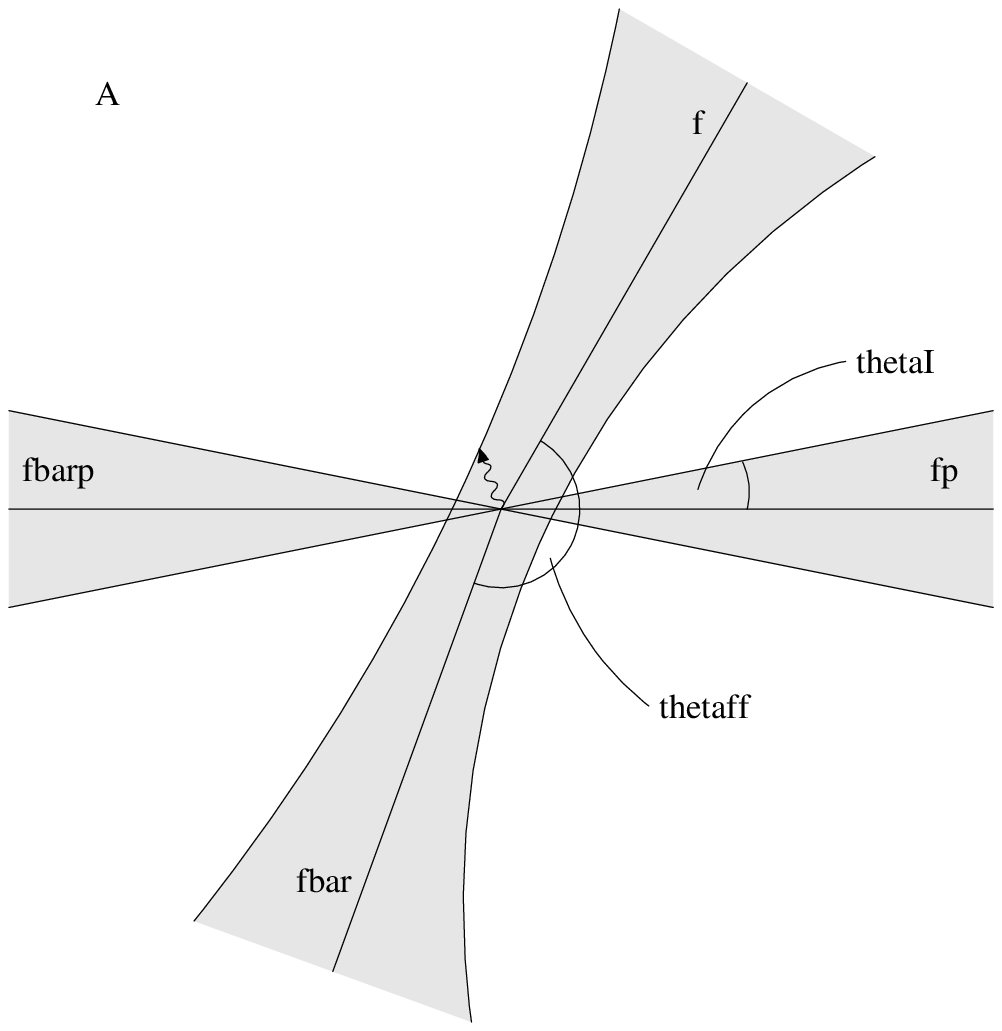}
 \hspace{10mm}
 \psfrag{fbar}{\small\hspace{-5mm}$\bar f$}
 \includegraphics[width=7cm]{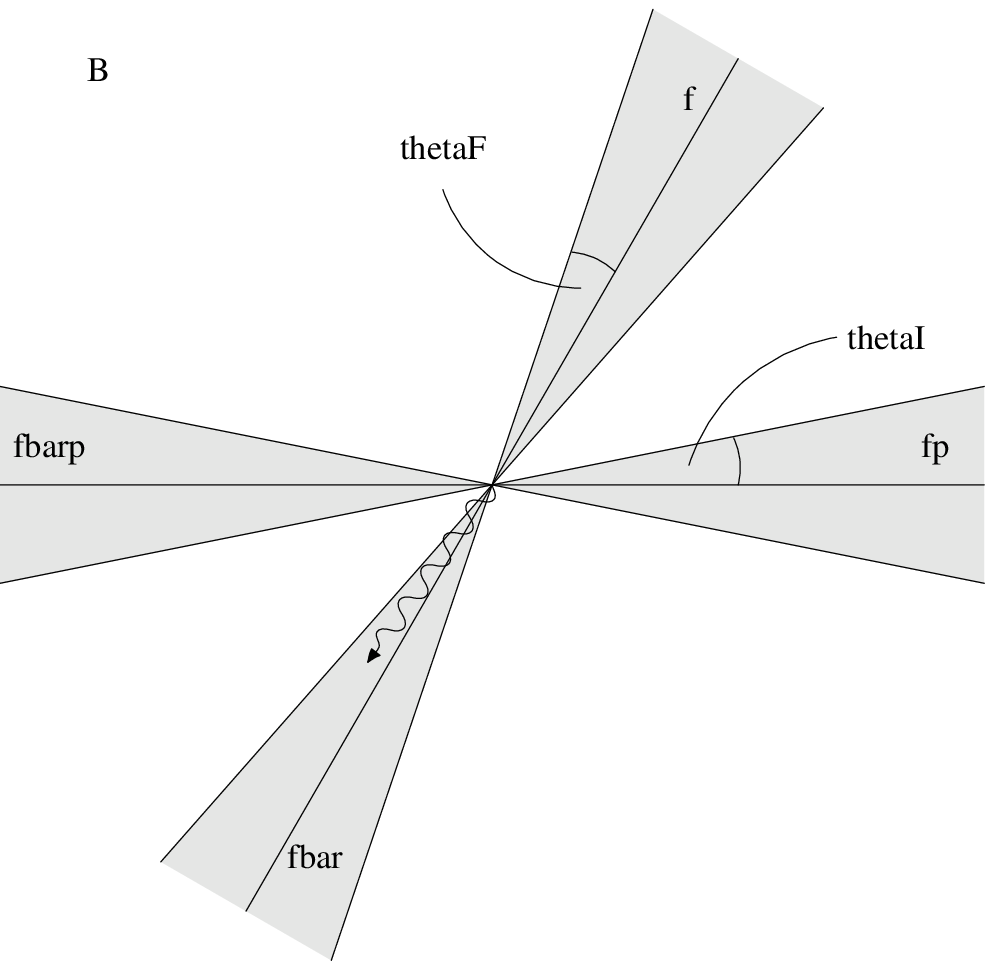}
}
\caption{\label{fig:scenarios} \small \textit{
Different restrictions on the real emission process. The momentum of the
undetected gauge boson (wavy line) is assumed to lie within the shaded
area. In Scenario~A (collinear and soft) we require the final state
fermions to be almost back-to-back or the emitted boson to lie within a
cone around the initial state fermions. In Scenario~B (collinear) the
emitted boson has to be within any of the cones around the fermions.}}}}
\end{figure}

The second new element of the four-fermion process consists in the fact
that we want to impose several phase space restrictions at once. In
particular we find it convenient to distinguish the following two
scenarios, which we will reconsider in our phenomenological analysis in
Section~\ref{sec:SM} (for an illustration of the scenarios
cf.~Figure~\ref{fig:scenarios}):  
\begin{itemize}
\item
In Scenario~A we combine virtual corrections with real gauge boson
radiation, if the final state fermions are almost back-to-back, with an
opening angle $\theta_{f\bar f}\geq\theta_{f\bar f}^c$, \emph{or} if the
emitted gauge boson is almost collinear to one of the initial state
fermions, i.e.~if $\theta_{f' b}\leq\theta_{I b}^c$  or $\theta_{\bar f'
  b}\leq\theta_{I b}^c$. Applying these phase space restrictions in
addition to the $Q^2$-cut discussed above, we find the logarithmic terms
to be 
\begin{align}
\sigma^{(R)}(z_c,\theta_{I b}^c,\theta_{f\bar f}^c) &\simeq
\frac{\alpha}{4\pi}
\left\{ 4 \bigg[ \ln^2 \frac{s}{M^2} - 3 \ln \frac{s}{M^2} \bigg]
\right. 
\no\\
&\hspace{14mm}\left.
+ 2 \left[ 1 + 4z_c + z_c^2 - 2\ln \frac{ z_c  }{ (1-z_c)^4} \right] \ln
\frac{s}{M^2} 
+ \ldots \right\} \sigma_B.
\label{eq:U1:fourferm:real:scenA}
\end{align}
In comparison with~(\ref{eq:U1:fourferm:real:Q2}) we see that the
additional phase space cuts $\theta_{I b}^c$ and $\theta_{f\bar f}^c$ do
not further modify the pattern of the Sudakov logarithms since all of
the singular regions (collinear to initial and final state and soft) are
covered in this scenario. We therefore expect a strong cancellation of
the virtual corrections even for tight cuts on $\theta_{I b}^c$ and
$\theta_{f\bar f}^c$.  
\item
In Scenario~B we require that the undetected vector boson is almost
collinear to one of the initial state fermions, i.e.~$\theta_{f'
b}\leq\theta_{I b}^c$ or $\theta_{\bar f' b}\leq\theta_{I b}^c$,
\emph{or} to one of the final state fermions, $\theta_{f b}\leq\theta_{F
  b}^c$ or $\theta_{\bar f b}\leq\theta_{F b}^c$. As we do not account
for soft radiation that escapes the four cones around the fermions, we
now expect a somewhat modified logarithmic structure and, consequently,
the compensation of the virtual corrections to be less
effective. Specifically, we now obtain  
\begin{align}
\sigma^{(R)}(z_c,\theta_{I b}^c,\theta_{F b}^c) &\simeq
\frac{\alpha}{4\pi}
\left\{ 4 \bigg[ \ln^2 \frac{s}{M^2} - 3 \ln \frac{s}{M^2} \bigg]
\right. 
\no\\
&\hspace{-5mm}\left.
+ 2 \left[ 1 + 4z_c + z_c^2 - 2\ln \frac{ z_c  }{ (1-z_c)^4}
-g(\theta_{I b}^c,\theta_{F b}^c) \right] \ln \frac{s}{M^2}
+ \ldots \right\} \sigma_B,
\label{eq:U1:fourferm:real:scenB}
\end{align}
with ($c_{I b}^c \equiv \cos \theta_{I b}^c$,$c_{Fb}^c \equiv \cos
\theta_{F b}^c$) 
\begin{align}
g(\theta_{I b}^c,\theta_{F b}^c) &=
\frac{c_{I b}^c(3+(c_{I b}^c)^2)}{2} \ln
\left(\frac{1+c_{Fb}^c}{1-c_{Fb}^c}\right) 
+ \frac{c_{Fb}^c(3+(c_{Fb}^c)^2)}{2} \ln \left(\frac{1+c_{I b}^c}{1-c_{I
      b}^c}\right) 
\no\\
&\hspace{5mm}
+\frac{3 c_{I b}^c c_{Fb}^c}{2} \left(2-(c_{I
    b}^c)^2-(c_{Fb}^c)^2\right). 
\end{align}
\end{itemize}

\begin{figure}[t!]
\centerline{\parbox{15.8cm}{
\centerline{
 \psfrag{Q2cut}{\hspace{-6mm}\footnotesize $\Delta\sigma(z_c)/\sigma_B$}
 \psfrag{sqrts}{\hspace{12mm}\scriptsize ${\sqrt{s}}$~[TeV]}
 \includegraphics[width=8.2cm]{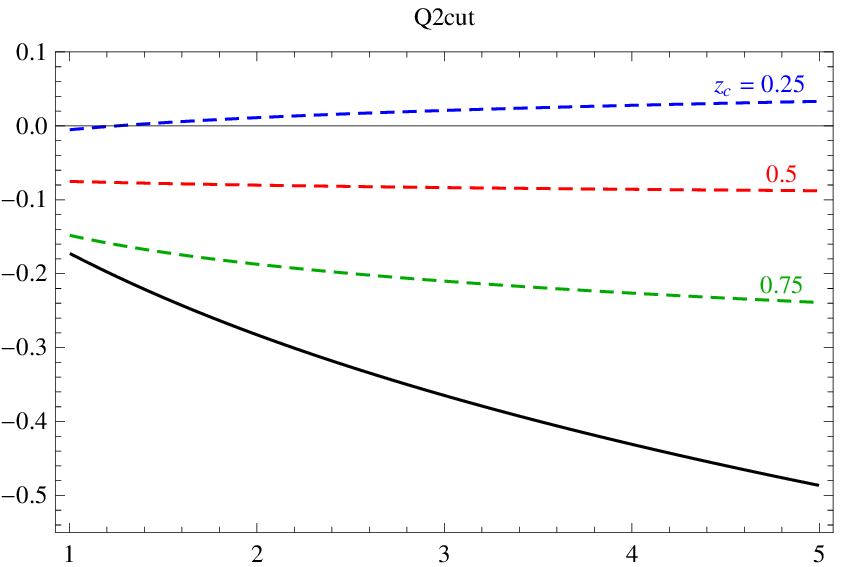}
}
\vspace{3mm}
\centerline{
 \psfrag{A}{\hspace{-23mm}\footnotesize
   Scenario~A:~$\Delta\sigma(z_c,\theta_{I b}^c,\theta_{f\bar
     f}^c)/\sigma_B$} 
 \psfrag{sqrts}{\hspace{12mm}\scriptsize ${\sqrt{s}}$~[TeV]}
 \includegraphics[width=8.2cm]{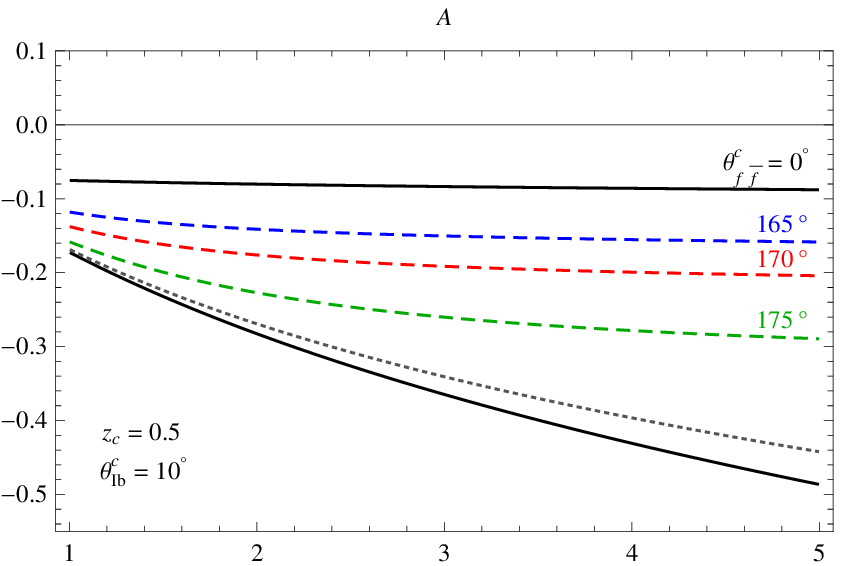}
}
\vspace{3mm}
\centerline{
 \psfrag{B}{\hspace{-23mm}\footnotesize
   Scenario~B:~$\Delta\sigma(z_c,\theta_{I b}^c,\theta_{F
     b}^c)/\sigma_B$} 
 \psfrag{sqrts}{\hspace{12mm}\scriptsize ${\sqrt{s}}$~[TeV]}
 \includegraphics[width=8.2cm]{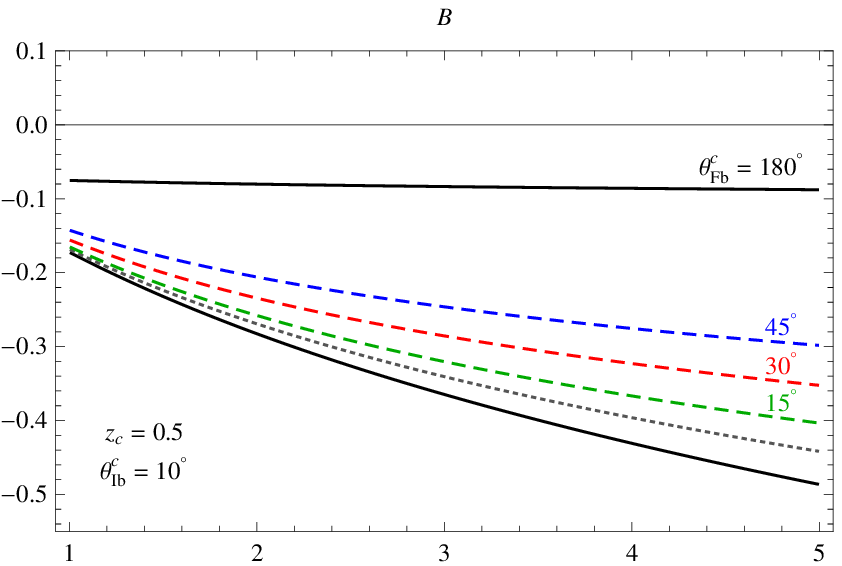}
}
\vspace{0mm}
\caption{\label{fig:U1:fourferm:cuts} \small \textit{
Relative NLO corrections to the four-fermion process in the abelian toy
theory as a function of the center of mass energy $\sqrt{s}$ in TeV. In
each plot the lower solid line represents the virtual correction (with
$M=80$~GeV, $\alpha=0.03$, $n_f=6$, $n_s=1$) and the dashed lines refer
to the sum $\Delta\sigma = \sigma^{(V)} + \sigma^{(R)}$ with different
restrictions on the real emission process. The individual dashed lines
(green/red/blue, from bottom to top in each plot) refer to
$z_c=0.75/0.5/0.25$ and no angular cut (top), $\theta_{f\bar f}^c =
175^\circ/170^\circ/165^\circ$~(middle) and $\theta_{F
b}^c=15^\circ/30^\circ/45^\circ$~(bottom). In the lower two plots we
fixed $z_c=0.5$ and $\theta_{I b}^c=10^\circ$. The dotted curves
indicate the contribution from initial state radiation (corresponding to
$\theta_{f\bar f}^c =180^\circ$ and $\theta_{F b}^c=0^\circ$,
respectively).}}}}  
\end{figure}

In Figure~\ref{fig:U1:fourferm:cuts} we illustrate these observations
quantitatively. First of all we note that the virtual corrections induce
a substantial Sudakov suppression in the TeV regime (in the abelian toy
theory with Standard Model inspired values $M=80$~GeV and
$\alpha=0.03$). Depending on the phase space cuts this suppression is
more or less compensated by the real emission process. In the upper plot
we illustrate the dependence on the cut on the invariant mass of the
fermion pair, $z_c=Q_c^2/s$, which is found to have a large impact on
the compensation (no angular cut has been applied so far). In the
remaining plots we fix $z_c=0.5$, i.e. the middle dashed line from the
first plot is the upper solid reference line for the other two
plots. We, moreover, impose a rather tight cut on initial state
radiation by setting $\theta_{I b}^c=10^\circ$ (the corresponding effect
is indicated by the dotted curves). From the middle plot it is evident
that the (unobserved) real radiation has a large impact in Scenario~A,
even when tight phase space cuts as $\theta_{f\bar f}^c=175^\circ$ are
applied. For more moderate cuts as $\theta_{f\bar f}^c=165^\circ$ the
virtual corrections are almost completely compensated in this setup. In
Scenario~B the compensation is found to be less effective. For moderate
cuts as $\theta_{F b}^c=30^\circ$ the virtual corrections are reduced,
for instance, from $-36\%$ to $-29\%$ at 3~TeV. This comparison
illustrates also quantitatively the importance of covering all of the
singular phase space regions.


\section{Bloch-Nordsieck violations}

\label{sec:BN}

The pattern of Sudakov logarithms is more complicated in non-abelian
gauge theories. The non-abelian group structure leads, in particular,
even for inclusive observables (with respect to  phase space) to a
mismatch between virtual and real Sudakov logarithms as long as one does
not sum over the non-abelian charges of the external particles. This
mismatch, commonly referred to as Bloch-Nordsieck (BN) violations, turns
out to be irrelevant in practical QCD applications, since the
confinement of the coloured partons into colour-neutral hadrons enforces
the summation (or average) over the colour charges. The spontaneous
breakdown of the electroweak interactions, however, allows to prepare
external states with definite weak isospin. Consequently, even inclusive
observables are affected by electroweak Sudakov
logarithms~\cite{CiaCiaCom}. 

In this section we reconsider the four-fermion process in a
spontaneously broken $SU(2)$ theory to study the structure and the
numerical impact of the BN violations. Whereas the gauge bosons
$W^{\pm,3}$ acquire a common mass $M$ in this non-abelian toy theory,
the fermions are again supposed to stay massless and to have a
vectorlike coupling to the gauge bosons. In the following we first
address the structure of the BN violations on the level of the total
cross section, then we switch to a quantitative analysis that accounts
for the various phase space restrictions that we introduced in the
previous section.

\subsection{Structure of Sudakov logarithms}
\label{sec:BN:logs}

The dynamical origin of Sudakov logarithms is well understood; they are
tied to the infra\-red structure of the theory and arise from collinear
or soft radiation of (almost) massless particles. Whereas previous
analyses have mainly focused on electroweak Sudakov logarithms from
virtual particle exchange
(cf.~e.g.~\cite{DegSir,KPS,KMPS,Jan,KMP,HKK1}), electroweak Sudakov
logarithms from real emission processes have received less attention so
far~\cite{CiaCiaCom,CiaCom2}. Let us therefore recall the origin and the
structure of the BN violations in the considered $SU(2)$ theory in some
detail. This will help us later to translate the results to Standard
Model processes.  

Let us start the discussion with the collinear approximation, which is
known to yield an universal radiation factor for each external
particle. This can be seen most easily in an axial gauge, where the
collinear logarithms stem from self energy insertions into external
lines. For the four-fermion process with generic isospin charges, $f_1
\bar{f}_2 \to f_3 \bar{f}_4$, the virtual collinear logarithms
associated with the outgoing fermion $f_3$ amount, for instance, to  
\begin{align}
\sigma^{(V, col~f_3)} &\simeq
- \frac{\alpha}{4\pi}
\bigg[ \ln^2 \frac{s}{M^2} - 3 \ln \frac{s}{M^2} \bigg]
(T^A T^A)_{f_3 f'} \;
\calC_B^{f_1 \bar{f}_2 \to f' \bar{f}_4} \;
\calC_B^{f_1 \bar{f}_2 \to f_3 \bar{f}_4} \;
\sigma_B^0,
\end{align}
where $T^A$ denotes a generator of the $SU(2)$ group and we made the
(real-valued) group structure of the Born amplitude, $\calA_B^{f_1
\bar{f}_2 \to f_3 \bar{f}_4} = \calC_B^{f_1 \bar{f}_2 \to f_3
\bar{f}_4}  \calA_B^0$, explicit ($\sigma_B^0\simeq 4\pi\alpha^2/3s$ is
the Born cross section of the abelian theory and a summation over $A$
and $f'$ is understood). In the collinear approximation we thus obtain a
Sudakov factor with negative weight for each external fermion, which is
to be multiplied with a Casimir factor $(T^A T^A)_{f_3 f'} = C_F\,
\delta_{f_3 f'}$. It is convenient to disentangle the contributions from
$W^3$ and $W^\pm$ exchange and to write the result as 
\begin{align}
\sigma^{(V, col~f_3)} &\simeq
 - \frac{\alpha}{4\pi}
\bigg[ \ln^2 \frac{s}{M^2} - 3 \ln \frac{s}{M^2} \bigg]
\Big( (t^3_{f_3})^2 + (t^\pm)^2 \Big) \; \sigma_B^{f_1 \bar{f}_2 \to f_3
  \bar{f}_4}, 
\label{eq:SU2:coll:V}
\end{align}
where $t^3_{f_3}$ denotes the isospin of the fermion $f_3$ and
$t^\pm=1/\sqrt{2}$ reflects the universal coupling to the ''charged
current''.  

In the same approximation Sudakov logarithms from real emission
processes can be derived on the basis of universal splitting
functions. We thus start in this case from the cross section
differential in $z = 1 + M^2/s - 2 E_W/\sqrt{s}$, where $E_W$ is the
energy of the emitted gauge boson in the center of mass frame. The
collinear logarithms associated with the outgoing fermion $f_3$ now
become  
\begin{align}
\frac{d\sigma^{(R, col~f_3)}}{dz} &\simeq
\frac{\alpha}{2\pi}
\bigg[ \frac{1+z^2}{1-z} \ln \frac{(1-z)^2s}{M^2} \bigg]
T^A_{f_3 f'} \; \calC_B^{f_1 \bar{f}_2 \to f' \bar{f}_4} \;
T^A_{f_3 f''} \; \calC_B^{f_1 \bar{f}_2 \to f'' \bar{f}_4} \;
\sigma_B^0.
\end{align}
Integrating this contribution in the kinematic limits $0\leq z \leq (1-
M/\sqrt{s})^2$ and disentangling again the contributions from $W^3$ and
$W^\pm$ emission, yields  
\begin{align}
\sigma^{(R, col~f_3)} &\simeq
\frac{\alpha}{4\pi}
\bigg[ \ln^2 \frac{s}{M^2} - 3 \ln \frac{s}{M^2} \bigg]
\Big( (t^3_{f_3})^2 \; \sigma_B^{f_1 \bar{f}_2 \to f_3 \bar{f}_4}
+ (t^\pm)^2 \; \sigma_B^{f_1 \bar{f}_2 \to f_3^\pm \bar{f}_4}
\Big),
\label{eq:SU2:coll:R}
\end{align}
where $f_3^\pm$ collectively denotes the isospin conjugate of the
fermion $f_3$, i.e.~$u^-= d$ and $d^+ = u$. Together with
(\ref{eq:SU2:coll:V}) we see that the Sudakov logarithms from $W^3$
exchange cancel between virtual and real corrections. This is, however,
different for $W^\pm$ exchange since the individual contributions
factorize to different Born cross sections.  

Let us briefly comment on the situation when the considered fermion is
in the initial state. The differential cross section contains in this
case an additional factor $1/z$, since the center of mass energy of the
hard subprocess has been lowered by the emission process. One may
further proceed along the lines of our explicit calculation in
(\ref{eq:U1:fourferm:real:plusdis}), which again yields a Sudakov factor
as in (\ref{eq:SU2:coll:R}) and a remnant collinear logarithm in the
plus-distribution which is left uncancelled.  

Soft gauge boson radiation induces further single logarithms. In
contrast to the collinear logarithms considered so far, the soft
logarithms are angular dependent and stem from interference effects. We
thus have to consider gauge boson exchange between pairs of
particles. The soft logarithms can be derived in the eikonal
approximation, which for an exchange between the incoming fermion $f_1$
and the outgoing fermion $f_3$ yields  
\begin{align}
\frac{d\sigma^{(V, soft~f_1f_3)}}{dt_{13}} &\simeq
-\frac{\alpha}{2\pi} \;
\ln^2 \frac{|t_{13}|}{M^2} \;
\frac{d\sigma_B^0}{dt_{13}} \;\,
T^A_{f' f_1} T^A_{f_3 f''} \;
\calC_B^{f' \bar{f}_2 \to f'' \bar{f}_4} \;
\calC_B^{f_1 \bar{f}_2 \to f_3 \bar{f}_4},
\end{align}
where $t_{13} = (p_{f_3} - p_{f_1})^2$. We next reshuffle the logarithm
according to  
\begin{align}
\ln^2 \frac{|t_{13}|}{M^2} =
\ln^2 \frac{s}{M^2} +
2\ln \frac{|t_{13}|}{s} \ln \frac{s}{M^2} +
\ln^2 \frac{|t_{13}|}{s},
\label{eq:SU2:soft:V:Logs}
\end{align}
and discard the double Sudakov logarithm since it originates from the
soft-collinear momentum region that we already accounted for in the
collinear approximation. The single angular-dependent logarithm in the
second term then leads to  
\begin{align}
\sigma^{(V, soft~f_1f_3)} &\simeq
-\frac{\alpha}{\pi} \;
N_{13} \;
\ln \frac{s}{M^2}
\bigg( t^3_{f_1} t^3_{f_3} \;
\sigma_B^{f_1 \bar{f}_2 \to f_3 \bar{f}_4}
+ (t^\pm)^2 \;
\calC_B^{f_1^\pm \bar{f}_2 \to f_3^\pm \bar{f}_4} \;
\calC_B^{f_1 \bar{f}_2 \to f_3 \bar{f}_4} \;
\sigma_B^0
\bigg),
\label{eq:SU2:soft:V}
\end{align}
where the prefactor $N_{13}$ encodes the angular integration,
\begin{align}
N_{13} =
\int dt_{13} \;
\ln \frac{|t_{13}|}{s}  \;
\frac{d\sigma_B^0/dt_{13}}{\sigma_B^0},
\end{align}
which, despite our assumption $s,|t|,|u|\gg M^2$, can be performed over
all angles in the logarithmic approximation. Let us add that the single
soft logarithms are absent if we pair two particles that are both in the
initial or final state, since $\ln^2 |s_{12}|/M^2 = \ln^2 |s_{34}|/M^2 =
\ln^2 s/M^2$ for the four-fermion process. The eikonal approximation
leads, moreover, to additional minus signs if we exchange incoming with
outgoing particles and fermions with antifermions.  

The corresponding logarithms from real emission processes can be derived
similarly within the eikonal approximation. For the same exchange
between the incoming fermion $f_1$ and the outgoing fermion $f_3$ we
obtain  
\begin{align}
\frac{d^2\sigma^{(R, soft~f_1f_3)}}{dt_{13}} &\simeq
\frac{\alpha}{2\pi} \;
\ln^2 \frac{|t_{13}|}{M^2} \;
\frac{d\sigma_B^0}{dt_{13}} \;\,
T^A_{f' f_1} \;
\calC_B^{f' \bar{f}_2 \to f_3 \bar{f}_4} \;
T^A_{f_3 f''} \;
\calC_B^{f_1 \bar{f}_2 \to f'' \bar{f}_4}.
\end{align}
Proceeding as before with (\ref{eq:SU2:soft:V:Logs}) and extracting the
contribution that encompasses the single soft logarithm, we get  
\begin{align}
\sigma^{(R, soft~f_1f_3)} &\simeq
\frac{\alpha}{\pi} \;
N_{13} \;
\ln \frac{s}{M^2}
\bigg( t^3_{f_1} t^3_{f_3} \;
\sigma_B^{f_1 \bar{f}_2 \to f_3 \bar{f}_4}
+ (t^\pm)^2 \;
\calC_B^{f_1^\pm \bar{f}_2 \to f_3 \bar{f}_4} \;
\calC_B^{f_1 \bar{f}_2 \to f_3^\mp \bar{f}_4}
\sigma_B^0
\bigg).
\end{align}
Together with (\ref{eq:SU2:soft:V}) we again see that the $W^3$
contribution cancels between virtual and real corrections, while the
$W^\pm$ contribution does not due to the modified group structure.  

In phenomenological applications one is often interested in observables
that are \emph{partly inclusive} in the non-abelian charges (e.g.~in
processes with light quarks in the final state). Let us therefore
briefly address the cross sections $\sigma_{u\bar u}$ and
$\sigma_{u\bar{d}}$, where the isospin charges of the initial state
particles have been fixed while the final state is considered to be
inclusive (for the neutral current we thus sum, for instance, over
$u'\bar u' W^3$, $u'\bar d' W^-$, etc.~where $u'/d'$ refer to a
different isospin doublet than $u/d$). As the BN violations from the
final state particles are washed out for these observables, the sum of
virtual and real corrections is free from angular-dependent
logarithms. We thus obtain a particularly simple
result~\cite{CiaCiaCom}, 
\begin{align}
\Delta \sigma_{u\bar u} &=
\sigma_{u\bar u}^{(V)} + \sigma_{u\bar u}^{(R)}
\simeq
\frac{\alpha}{4\pi}
\bigg[ \ln^2 \frac{s}{M^2} - 3 \ln \frac{s}{M^2} \bigg] (t^\pm)^2  \;
\Big( \sigma_{u \bar d}^B + \sigma_{d \bar u}^B -2 \sigma_{u \bar u}^B\big),
\no\\
\Delta \sigma_{u\bar d} &=
\sigma_{u\bar d}^{(V)} + \sigma_{u\bar d}^{(R)}
\simeq
\frac{\alpha}{4\pi}
\bigg[ \ln^2 \frac{s}{M^2} - 3 \ln \frac{s}{M^2} \bigg] (t^\pm)^2  \;
\Big( \sigma_{u \bar u}^B + \sigma_{d \bar d}^B -2 \sigma_{u \bar d}^B\big),\
\label{eq:SU2:partialsums:general}
\end{align}
where we suppressed the logarithms from initial state radiation that are
not supposed to cancel (plus-distributions). Given the Born relations
$\sigma_{u \bar d}^B=\sigma_{d \bar u}^B=2\sigma_{u \bar u}^B=2
\sigma_{d \bar d}^B$, we get  
\begin{align}
\Delta \sigma_{u\bar u}
= - \Delta \sigma_{u\bar d} &\simeq
\frac{\alpha}{4\pi}
\bigg[ \ln^2 \frac{s}{M^2} - 3 \ln \frac{s}{M^2} \bigg] (t^\pm)^2  \;
\sigma_{u \bar d}^B.
\label{eq:SU2:partialsums}
\end{align}
On the basis of the BN violations, we thus expect an overcompensation of
the virtual corrections for the inclusive neutral current process, while
the real radiation is expected to be less important for the charged
current.

\subsection{NLO calculation with phase space cuts}
\label{sec:BN:NLO}

Let us now investigate the numerical impact of the BN violations and
their interplay with the phase space cuts. In contrast to the
considerations from the previous section, we now consider the full NLO
calculation accounting for double and single Sudakov logarithms as well
as constant terms. Power-suppressed terms of $\mathcal{O}(M^2/s)$, on
the other hand, will be neglected.  

In this approximation the one-loop virtual corrections to the
$s$-channel four-fermion process $f_1 \bar{f}_2 \to f_3 \bar{f}_4$ 
become 
\begin{align}
\sigma^{(V)}
&\simeq \frac{\alpha}{4\pi}
\left \{ - 4 C_F \bigg[ \ln^2 \frac{s}{M^2} - 3 \ln \frac{s}{M^2} \bigg]
+ \frac{13}{3} C_A \ln \frac{s}{M^2}
+ \left( \frac{4\pi^2}{3} - 14 \right) C_F \right.
\no\\
&\hspace{14mm}
\left.
+ \left(\frac{259}{18} -2 \pi^2 \right) C_A
-\frac{40}{9}  T_F n_f
-\frac{8}{9} n_s
\right\} \sigma_B^{f_1 \bar{f}_2 \to f_3 \bar{f}_4},
\label{eq:SU2:fourfermion:V}
\end{align}
which implies the same \emph{relative} correction for charged and
neutral current processes. The Sudakov logarithms in
(\ref{eq:SU2:fourfermion:V}) have a simple interpretation in terms of
our formal analysis from the previous section: the Sudakov factors $\sim
C_F$ stem from the collinear approximation (\ref{eq:SU2:coll:V}), while
the soft logarithms $\sim C_A$ result from the various pairings
(\ref{eq:SU2:soft:V}) of external particles\footnote{The soft logarithms
  were absent in the abelian theory,  cf.~(\ref{eq:U1:fourfermion:V}),
since the sum of all pairings led to an exact cancellation in this
case.}.  

\begin{figure}[t!]
\centerline{\parbox{15.8cm}{
\centerline{
 \psfrag{uuA}{\hspace{-16mm}\footnotesize Scenario~A:~$(\Delta
   \sigma/\sigma_B)_{u \bar u}$} 
 \includegraphics[width=7.6cm]{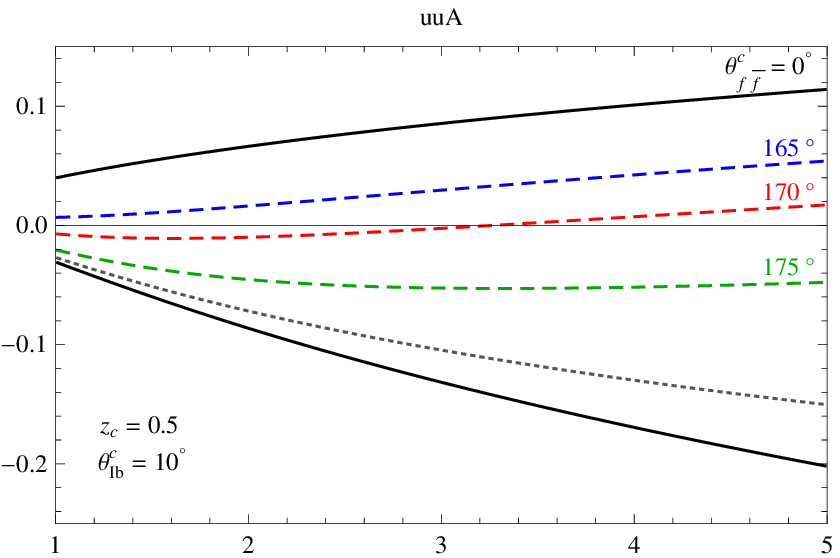}
\hspace{0.5cm}
 \psfrag{uuB}{\hspace{-16mm}\footnotesize Scenario~B:~$(\Delta
   \sigma/\sigma_B)_{u \bar u}$} 
 \includegraphics[width=7.6cm]{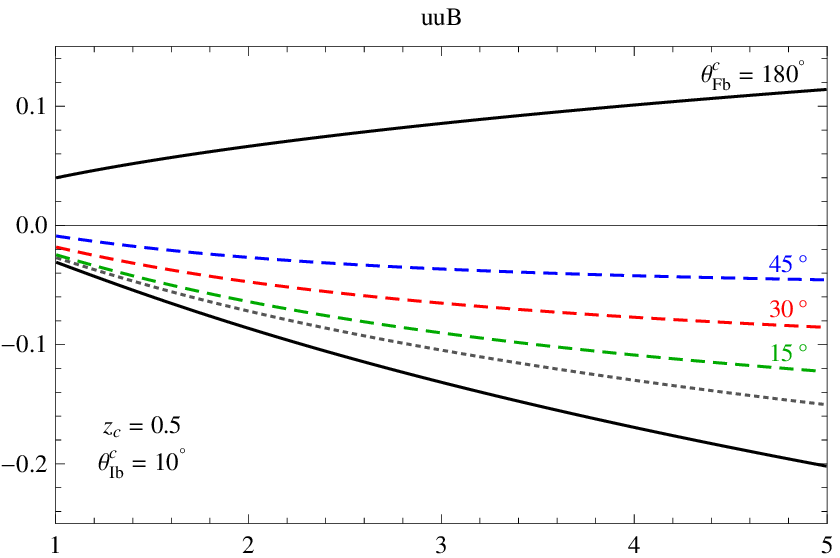}
}
\vspace{8mm}
\centerline{
 \psfrag{udA}{\hspace{-16mm}\footnotesize Scenario~A:~$(\Delta
   \sigma/\sigma_B)_{u \bar d}$} 
 \includegraphics[width=7.6cm]{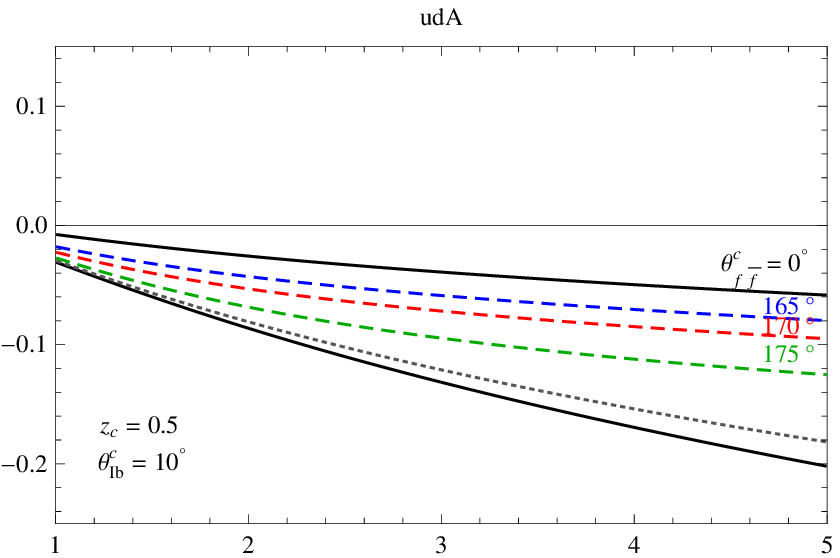}
\hspace{0.5cm}
 \psfrag{udB}{\hspace{-16mm}\footnotesize Scenario~B:~$(\Delta
   \sigma/\sigma_B)_{u \bar d}$} 
 \includegraphics[width=7.6cm]{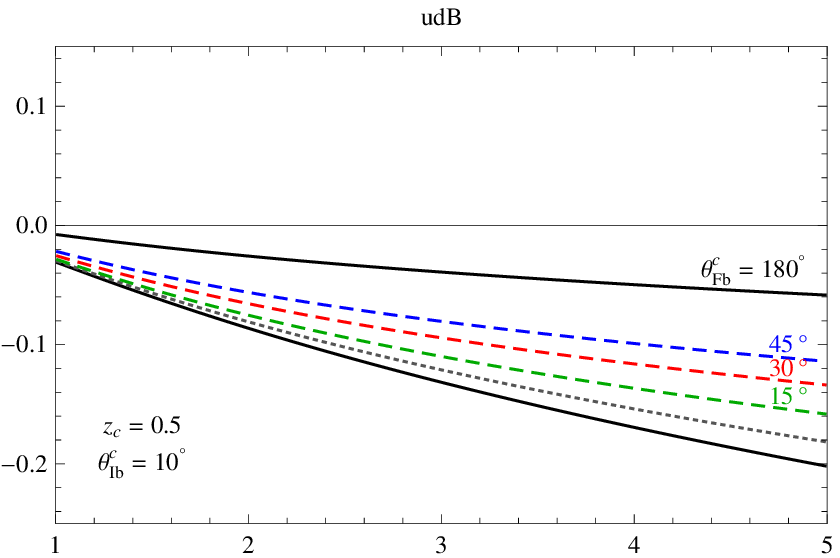}
}
\vspace{2mm}
\caption{\label{fig:SU2:fourferm:cuts} \small \textit{
Relative NLO corrections to the four-fermion process in a spontaneously
broken $SU(2)$ theory with neutral (upper plots) and charged (lower
plots) initial states (notation and numerical input values from
Figure~\ref{fig:U1:fourferm:cuts}).}}}}  
\end{figure}

In the next step we include a certain amount of (unobservable) real
gauge boson radiation according to our scenarios from
Figure~\ref{fig:scenarios}. In view of the phenomenological applications
from Section~\ref{sec:SM}, we will concentrate on the semi-inclusive
cross sections $\sigma_{u\bar u}$ and $\sigma_{u\bar{d}}$ that we
introduced at the end of the previous section. For both sets of cuts the
relative NLO corrections are shown in
Figure~\ref{fig:SU2:fourferm:cuts}. First of all we note that the
Sudakov suppression is somewhat less pronounced in the non-abelian toy
theory due to the prefactor $C_F=3/4$ multiplying the Sudakov factors
and the impact of the soft logarithms which happen to contribute with
opposite sign. Still, the one-loop virtual corrections induce a
$10$-$20\%$ suppression in the TeV regime.  Comparing left and right
plots, we recognize the qualitative difference between the two scenarios
that we worked out in detail for the abelian theory\footnote{Closer
inspection reveals that the dependence on the cutoff $z_c$,
cf.~(\ref{eq:U1:fourferm:real:Q2}) and
(\ref{eq:U1:fourferm:real:scenA}), is also present in the non-abelian
theory, while the soft logarithms only slightly modify the angular
dependence $g(\theta_{Ib}^c,\theta_{Fb}^c)$ in
(\ref{eq:U1:fourferm:real:scenB}).}. The abelian picture is, moreover,
significantly modified by the BN violations. This is in particular true,
when soft and collinear radiation is allowed by the phase space cuts and
the compensation is further favoured by the BN violations (upper left
plot). Here even for a tight cut like $\theta_{f\bar f}^c=175^\circ$
almost complete compensation of the Sudakov suppression is observed,
while for a more moderate cut as $\theta_{f\bar f}^c=165^\circ$ we find
an overcompensation of the virtual corrections. It is also interesting
to compare the upper right plot (neutral initial state, collinear
radiation included) with the lower left plot (charged initial state,
soft and collinear radiation included), where one reads off that the
differences between the two scenarios can, at least to some extent, be
washed out by the BN violations.   

In total we find that the large negative corrections from virtual gauge
boson exchange can be partially compensated if real radiation is
included. The details of this compensation mechanism depend on the
isospin configuration of initial and final state particles and on the
particular phase space cuts that constrain the real radiation.


\section{Electroweak Sudakov corrections}

\label{sec:SM}

Having exemplified the basic concepts behind the compensation of virtual
Sudakov corrections in spontaneously broken gauge theories, we may now
translate our observations to Standard Model processes. Let us emphasize
that this discussion will be, necessarily, of qualitative nature. The
efficiency of the cuts on gauge boson radiation will depend on the
details of the specific process as, for instance, the fermionic initial
and final state (leptons or quark jets), the charge of the radiated
gauge boson ($Z$ or $W$) and its decay mode. Restricting the discussion
for example to $e^+e^-$ colliders, it is plausible that $\mu^+\mu^-$
will constitute a ``clean'' final state and gauge boson radiation can be
rejected. For quark-antiquark final states, on the other hand, collinear
energetic gauge bosons decaying hadronically may well be masked by the
quark jets. In contrast, it is plausible that leptonically decaying $Z$
bosons can be separated from the background. Soft gauge bosons emitted
under large angles will again lead to different signatures. The
efficiency for detection of gauge boson radiation in hadronic collisions
will again be different. 

With the kinematics at hadron colliders being less constrained as a
consequence of the convolution with parton distribution functions, the
following discussion will be restricted to the simpler case of
electron-positron collisions while hadron collisions will be treated at
a later point~\cite{BKR}. Since one may expect that reactions with
leptonic final states will be fairly clean and not ``contaminated'' by
gauge boson radiation, we will concentrate on the process $e^+e^-\to
q\bar q$ (where $q=u,d$). This allows us to study both of the aspects
that we discussed in the previous sections in a realistic environment:
First, the process is affected by BN violations since the isospin
charges of the initial state fermions are singled out and, second, soft
and/or collinear gauge boson emission may not always be easily resolved
in an experimental setup due to the hadronic signature of the process.

\subsection{Bloch-Nordsieck violations}

Let us first address the BN violations of the current process without
restrictions on the gauge boson kinematics. From our analysis in
Section~\ref{sec:BN:logs} we deduce that $Z$ boson (and photon) emission
is irrelevant in this context. Moreover, as we sum over the quark
flavours $u$ and $d$ in the final state, the respective BN violations
are washed out. We are thus left with the first of the equations in
(\ref{eq:SU2:partialsums:general}).  

In the chiral electroweak theory we have, in addition, to specify the
helicity structure of the process. For left-handed leptons in the
initial and left or right-handed quarks in the final state, we obtain  
\begin{align}
\Delta \sigma_{e^-e^+}^{LL} &\simeq
\frac{\alpha}{4\pi s_w^2}
\bigg[ \ln^2 \frac{s}{M_W^2} - 3 \ln \frac{s}{M_W^2} \bigg] \;
\Big( \sigma_{e^- \bar \nu}^{B,LL} - \sigma_{e^- e^+}^{B,LL}\big),
\no\\
\Delta \sigma_{e^-e^+}^{LR} &\simeq
\frac{\alpha}{4\pi s^2_w}
\bigg[ \ln^2 \frac{s}{M_W^2} - 3 \ln \frac{s}{M_W^2} \bigg] \;
\Big( - \sigma_{e^- e^+}^{B,LR}\big),
\end{align}
respectively, where $s_w^2=\sin^2 \theta_w\simeq0.231$ with $\theta_w$
being the weak mixing angle. As $\sigma_{e^- \bar \nu}^{B,LL} \simeq
1.98~\sigma_{e^- e^+}^{B,LL}$ the purely left-handed component shows an
overcompensation of the virtual corrections, similar to the neutral
current process from the $SU(2)$ theory in Section~\ref{sec:BN}. The
situation is, however, reversed for right-handed quarks since the
corresponding charged current process is forbidden at Born level,
$\sigma_{e^- \bar \nu}^{B,LR}=0$. The BN violations are, moreover,
absent for right-handed leptons in the initial state,  
\begin{align}
\Delta \sigma_{e^-e^+}^{RL} &=
\Delta \sigma_{e^-e^+}^{RR} \simeq
0.
\end{align}
The cancellation of the logarithmic $W$-corrections for left-handed
quarks is a consequence of the fact that we sum over quark flavours in
the final state, while the BN violations are of course absent for the
purely abelian right-handed component.  

We thus see that the chiral coupling of the $W$-bosons induces distinct
patterns of BN violations that depend on the helicities of the external
particles. In the present case this leads, in particular, to interesting
effects for polarized lepton beams. For unpolarized beams, on the other
hand, the purely left-handed component is expected to dominate the
pattern since $\sigma_{e^-e^+}^{B,LL} \simeq 2.52~\sigma_{e^-e^+}^{B,RR}
\simeq 10.1~\sigma_{e^-e^+}^{B,LR} \simeq 25.2~\sigma_{e^-e^+}^{B,RL}$
at Born level.

\subsection{Numerical analysis}

Let us first inspect the virtual corrections to the current process in
some detail. Neglecting power-suppressed terms of
$\mathcal{O}(M_{W,Z}^2/s)$ and summing over the helicities of incoming
and outgoing fermions, the one-loop electroweak corrections for the
annihilation into (massless) up- and down-type quarks can be written as
\begin{align}
\sigma^{(V)}_{e^-e^+\to u \bar u}
&\simeq \frac{\alpha}{4\pi s_w^2}
\left \{ - 1.28 \bigg[ \ln^2 \frac{s}{M_W^2} - 3 \ln \frac{s}{M_W^2}
  \bigg] 
+ 1.43 \ln \frac{s}{M_W^2} \right.
\no\\
&\hspace{26.3mm}
\left.
- 0.39 \bigg[ \ln^2 \frac{s}{M_Z^2} - 3 \ln \frac{s}{M_Z^2} \bigg]
- 1.12 \ln \frac{s}{M_Z^2}
-8.36
\right\} \sigma^{B}_{e^-e^+\to u \bar u},
\no\\
\sigma^{(V)}_{e^-e^+\to d \bar d}
&\simeq \frac{\alpha}{4\pi s_w^2}
\left \{ - 1.62 \bigg[ \ln^2 \frac{s}{M_W^2} - 3 \ln \frac{s}{M_W^2}
  \bigg] 
+ 12.57 \ln \frac{s}{M_W^2} \right.
\no\\
&\hspace{26.3mm}
\left.
- 0.56 \bigg[ \ln^2 \frac{s}{M_Z^2} - 3 \ln \frac{s}{M_Z^2} \bigg]
+ 1.48 \ln \frac{s}{M_Z^2}
- 34.02
\right\} \sigma^{B}_{e^-e^+\to d \bar d},
\label{eq:SM:fourfermion:V}
\end{align}
where we distinguished between Sudakov factors, which encode the
collinear logarithms, and single soft logarithms from $W$- and $Z$-boson
exchange. Note that the soft logarithms come in the latter case with a
large positive coefficient, which significantly reduces the Sudakov
suppression in the few TeV region. The relative corrections to the
inclusive process $e^+e^-\to q\bar q$ amount, for instance, to
$-2.7\%$~$(-6.6\%)$ at $\sqrt{s}=1$~TeV ($2$~TeV), respectively.  

One comment is in order concerning our treatment of QED divergences. As
our prior interest are ''genuine'' electroweak effects from $W$- and
$Z$-boson emission, we will disregard Sudakov effects of pure QED
nature. In other words we do \emph{not} include real photon emission in
our analysis, but rather subtract the QED divergences, which we
regularized with a photon mass, from the virtual corrections (they have
already been omitted in (\ref{eq:SM:fourfermion:V})). In order to obtain
a physical cross section, our results thus have to be supplemented by a
standard QED correction factor that depends on fermion masses and on
specific cuts that constrain the soft photon emission, but is
independent of $M_{W,Z}$. For the process under consideration this is a
gauge invariant separation.  

\begin{figure}[t!]
\centerline{\parbox{15.8cm}{
\centerline{
 \psfrag{A}{\hspace{-18mm}\footnotesize
   Scenario~A:~$(\Delta\sigma/\sigma_B)_{e^-e^+}$} 
 \psfrag{B}{\hspace{-18mm}\footnotesize
   Scenario~B:~$(\Delta\sigma/\sigma_B)_{e^-e^+}$} 
 \includegraphics[width=7.4cm]{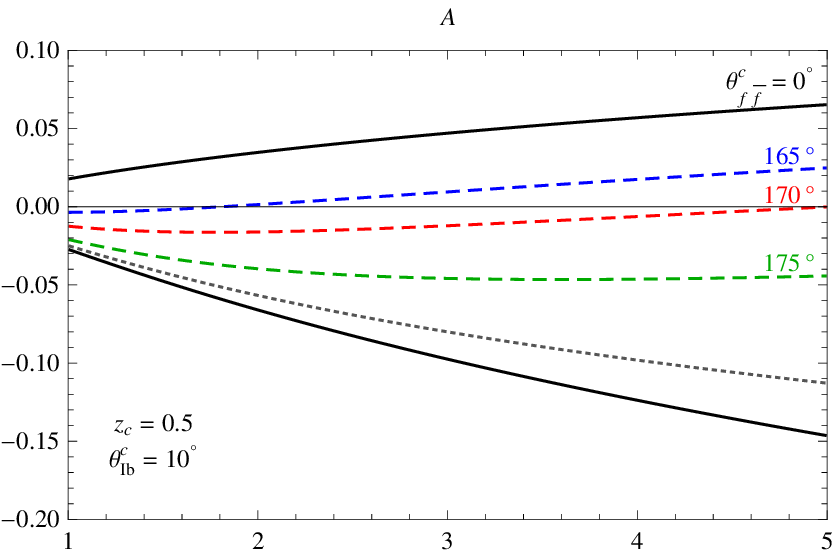}
\hspace{0.5cm}
 \includegraphics[width=7.4cm]{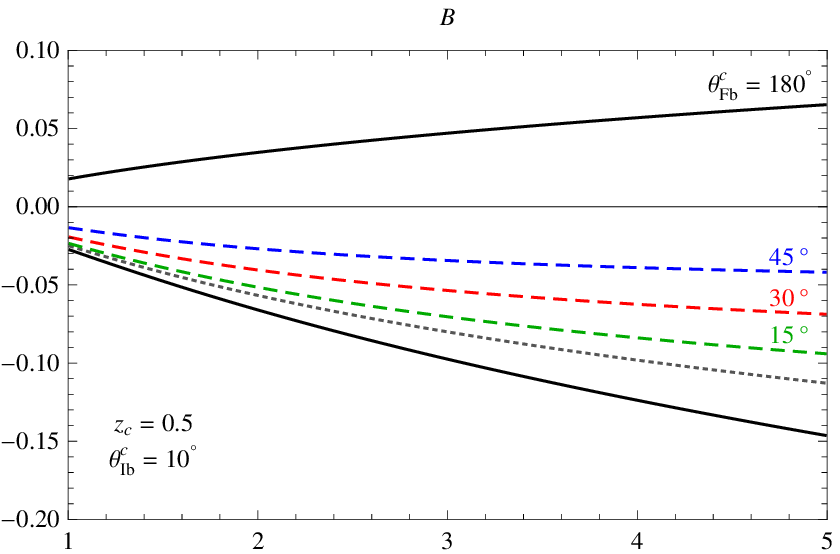}
}
\vspace{8mm}
\centerline{
 \psfrag{A}{\hspace{-18mm}\footnotesize
   Scenario~A:~$(\Delta\sigma/\sigma_B)_{e^-e^+}$} 
 \psfrag{B}{\hspace{-18mm}\footnotesize
   Scenario~B:~$(\Delta\sigma/\sigma_B)_{e^-e^+}$} 
 \includegraphics[width=7.4cm]{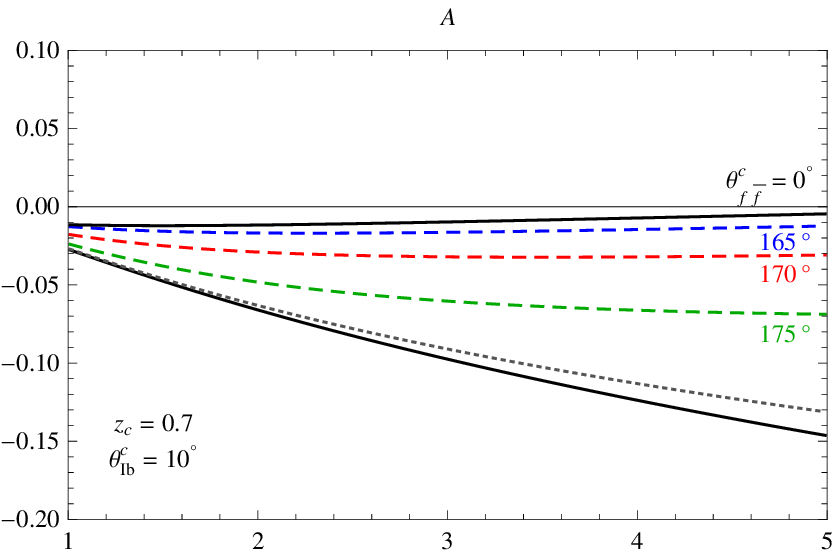}
\hspace{0.5cm}
\includegraphics[width=7.4cm]{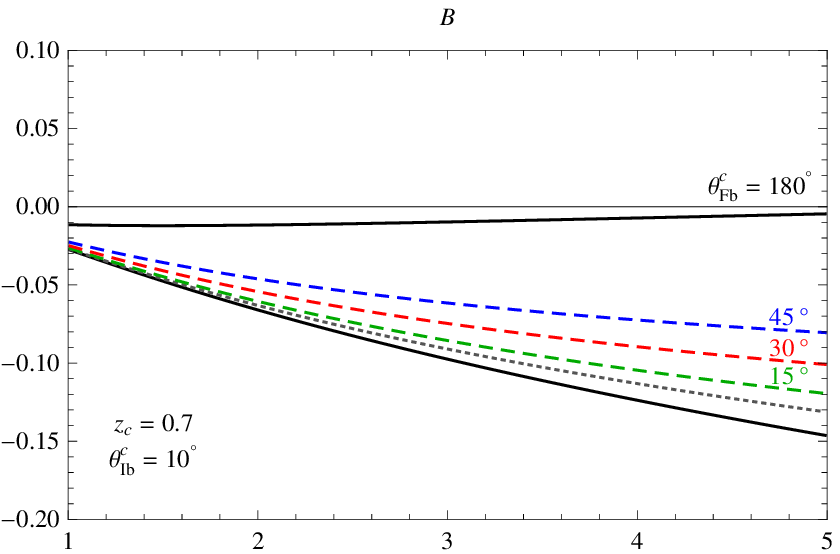}
}
\caption{\label{fig:SM:fourferm:cuts} \small \textit{
Relative NLO electroweak corrections to $e^+e^-\to q\bar q$ as a
function of the center of mass energy in TeV. In each plot the lower
solid line represents the virtual correction (with $\alpha=1/128$ and
$s_w^2=0.231$) and the dashed lines refer to the sum $\Delta\sigma =
\sigma^{(V)} + \sigma^{(R)}$ with different restrictions on the real
emission process (according to the scenarios from
Figure~\ref{fig:scenarios}). The individual dashed lines
(green/red/blue, from bottom to top in each plot) refer to
$\theta_{f\bar f}^c = 175^\circ/170^\circ/165^\circ$ (left) and
$\theta_{F b}^c=15^\circ/30^\circ/45^\circ$~(right). We further set
$z_c=0.5/0.7$ in the upper/lower plots and $\theta_{I
b}^c=10^\circ$. The dotted curves indicate the contribution from initial
state radiation (corresponding to $\theta_{f\bar f}^c =180^\circ$ and
$\theta_{F b}^c=0^\circ$, respectively).}}}} 
\end{figure}

Let us now turn to real $W$- and $Z$-boson radiation. Focusing again on
the process with unpolarized leptons in the initial state and summing
over the quark species and polarizations in the final state, we 
illustrate the size of the BN violations in the upper plots from 
Figure~\ref{fig:SM:fourferm:cuts} (adopting the same conventions as in
Figure~\ref{fig:SU2:fourferm:cuts}).  As the process is dominated by the
purely left-handed component, we essentially recover the pattern of the
neutral current process from the $SU(2)$ theory, cf.~the upper plots
from Figure~\ref{fig:SU2:fourferm:cuts}. Our default choice of phase
space cuts ($z_c=Q_c^2/s =0.5, \theta_{I b}^c=10^\circ$) may, however,
not be quite realistic for the considered process.  As an alternative we
therefore show the result for a more restrictive cut on initial state
radiation ($z_c=0.7$, $\theta_{I b}^c=10^\circ$) in the lower plots from
Figure~\ref{fig:SM:fourferm:cuts}. In other words we demand that the
quark-antiquark pair (or rather their associated jets) carries at least
about $84\%$ of the beam energy and we assume that $W$- and $Z$-bosons,
that are emitted into the extreme forward direction, cannot be resolved
for angles $\theta_{Ib}\leq10^\circ$ (which corresponds to a
pseudorapidity cut $|\eta|\geq2.4$). We may then investigate the impact
from soft and (final state) collinear $W$- and $Z$-boson radiation by
varying the parameters $\theta_{f\bar f}^c$ and $\theta_{Fb}^c$.  

From the lower plots in Figure~\ref{fig:SM:fourferm:cuts} we read off
that real gauge boson radiation becomes numerically relevant in the few
TeV region only if some fraction of collinear \emph{and} soft $W$- and
$Z$-bosons escapes experimental detection (Scenario A). For reasonable
values of phase space cuts as $\theta_{Fb}^c=15^\circ$ the Sudakov
suppression is, for instance, only marginally reduced in Scenario~B from
$-2.7\%$~$(-6.6\%)$ to $-2.6\%$~$(-6.1\%)$ at $\sqrt{s}=1$~TeV
($2$~TeV). In contrast to this the impact from real radiation is much
more pronounced if some soft non-collinear radiation is accounted for. 
In Scenario~A the Sudakov suppression is, for instance, reduced to 
$-1.8\%$~$(-2.9\%)$ for $\theta_{f\bar f}^c=170^\circ$.


\section{Conclusions}

\label{sec:conclusion}

The purpose of our work was to investigate to which extent real gauge
boson radiation can compensate the characteristic negative virtual
corrections that arise in high-energy reactions. As the latter are
driven by Sudakov logarithms, the compensation mechanism depends
obviously on the amount of soft and collinear radiation that is allowed
by the phase space cuts. A second interesting element in this context is
the mismatch of logarithmically enhanced terms in spontaneously broken
gauge theories, a phenomenon called Bloch-Nordsieck violations. 

In order to address these issues separately, we subsequently studied a
massive abelian gauge theory, a spontaneously broken $SU(2)$ theory and
the electroweak Standard Model. We derived analytical results for some
exemplary (and idealized) cuts, which facilitate the qualitative
understanding of the compensation mechanism. In our numerical analysis
we found remarkable differences for cuts which cover all of the singular
regions (collinear \emph{and} soft) and those that include them only
partially (collinear \emph{or} soft).  

The factorization of soft and collinear singularities can be exploited
to compute the Bloch-Nordsieck violations for inclusive cross sections
on a process-independent basis. Depending on the non-abelian charges of
the external particles, the Bloch-Nordsieck violations can lead to a
partial cancellation or to an overcompensation of the virtual
corrections. We argued that this can to some extent wash out the
qualitative differences of the phase space cuts.  

We, in particular, tried to understand to which extent electroweak
Sudakov corrections are affected by these issues. To this end we
discussed the case of electron-positron annihilation into $u\bar u$ and
$d\bar d$ quarks in more detail. We performed an explicit NLO
calculation and investigated the impact from unobservable $W$- and
$Z$-boson radiation. While the Sudakov suppression is not particularly
pronounced for this specific process, it allowed us to study the
compensation mechanism in a realistic environment. We found that real
radiation becomes numerically relevant for this process only if a
fraction of collinear \emph{and} soft $W$- and $Z$-bosons escapes
experimental detection. 

From the phenomenological point of view real $W$- and $Z$-boson
radiation is certainly more important for hadron collider
processes. Current hadron colliders are on the eve of probing the
multi-TeV region in which the Sudakov effects become more
pronounced. The hadronic environment makes, moreover, the discrimination
of real gauge boson radiation much more challenging. Typical observables
at hadron colliders are actually largely inclusive; the Drell-Yan
process allows, for instance, for an arbitrary number of $W$- and
$Z$-bosons decaying hadronically. We plan to extent the presented
analysis to hadron collider processes in a future
publication~\cite{BKR}.


\subsection*{Acknowledgements}


This work was supported by BMBF grant 05H09VKE, the
Sonderforschungsbereich/ 
Transregio 9 and the Graduiertenkolleg "Hochenergiephysik und
Teilchenastrophysik".


\end{document}